\documentclass[11pt]{article}
\usepackage{amsfonts}

\usepackage{amssymb}
\usepackage{latexsym}
\usepackage{graphicx}
\usepackage[english]{babel}

\begin{document}
\input epsf

\makeatletter
\@addtoreset{equation}{section}
\makeatother


 \begin{center}
{\LARGE The information paradox: conflicts and resolutions\footnote{(Expanded version of) proceedings for Lepton-Photon 2011.}}
\\
\vspace{18mm}
{\bf    Samir D. Mathur
}
\vspace{8mm}

\vspace{8mm}

Department of Physics,\\ The Ohio State University,\\ Columbus,
OH 43210, USA\\ 
\vskip 2 mm
mathur@mps.ohio-state.edu\\
\vspace{10mm}

\end{center}

\thispagestyle{empty}

\def\p{\partial}
\def\h{{1\over 2}}
\def\be{\begin{equation}}
\def\bea{\begin{eqnarray}}
\def\ee{\end{equation}}
\def\eea{\end{eqnarray}}
\def\d{\partial}
\def\la{\lambda}
\def\eps{\epsilon}
\def\bb{\bigskip}
\def\mm{\medskip}
\newcommand{\dm}{\begin{displaymath}}
\newcommand{\edm}{\end{displaymath}}
\renewcommand{\b}{\tilde{B}}
\newcommand{\gm}{\Gamma}
\newcommand{\ac}[2]{\ensuremath{\{ #1, #2 \}}}
\renewcommand{\ell}{l}
\newcommand{\z}{\ell}
\newcommand{\newsection}[1]{\section{#1} \setcounter{equation}{0}}
\def\bb{$\bullet$}
\def\Qbar{{\bar Q}_1}
\def\QPbar{{\bar Q}_p}
\def\q{\quad}
\def\bn{B_\circ}
\def\sq{{1\over \sqrt{2}}}
\def\z{|0\rangle}
\def\o{|1\rangle}
\def\sqi{{1\over \sqrt{2}}}

\let\a=\alpha \let\b=\beta \let\g=\gamma \let\d=\delta \let\e=\epsilon
\let\c=\chi \let\th=\theta  \let\k=\kappa
\let\l=\lambda \let\m=\mu \let\n=\nu \let\x=\xi \let\r=\rho
\let\s=\sigma \let\t=\tau
\let\vp=\varphi \let\vep=\varepsilon
\let\w=\omega      \let\G=\Gamma \let\D=\Delta \let\Th=\Theta
                     \let\P=\Pi \let\S=\Sigma

\def\h{{1\over 2}}
\def\t{\tilde}
\def\r{\rightarrow}
\def\nn{\nonumber\\}
\let\bm=\bibitem
\def\Kt{{\tilde K}}
\def\b{\bigskip}

\let\p=\partial
\def\u{\uparrow}
\def\d{\downarrow}

\begin{abstract}

Many relativists have been long convinced that black hole evaporation leads to information loss or remnants. String theorists have however not been too worried about the issue, largely due to a belief that the Hawking argument for information loss is flawed in its details. A recently derived  inequality shows that the Hawking argument for black holes with horizon can in fact be made rigorous. What happens instead is that in string theory black hole microstates have no horizons. Thus the evolution of radiation quanta with $E\sim kT$ is modified by order unity at the horizon, and we resolve the information paradox. We discuss how it is still possible for $E\gg kT$ objects to see an approximate black hole like geometry. We also note some possible implications of this physics for the early Universe.

\end{abstract}
\vskip 1.0 true in

\newpage
\renewcommand{\theequation}{\arabic{section}.\arabic{equation}}

\def\p{\partial}
\def\r{\rightarrow}
\def\h{{1\over 2}}
\def\b{\bigskip}

\def\nn{\nonumber\\ }

\section{Introduction}
\label{intr}\setcounter{equation}{0}

The quantum theory of black holes involves all three of the fundamental constants of nature -- $c, \hbar, G$. So we can hope that studying black holes will lead us towards an ultimate theory of all physics.

But in 1974 Stephen Hawking showed that what we get instead is a {\it contradiction} \cite{hawking}. Suppose a  star collapses to make a black hole with the structure expected from general relativity. Hawking showed that the natural evolution of quantum mechanics  will  take the initial wavefunction of the star to a state that {\it cannot} be  described by a  wavefunction. This conflict between general relativity and quantum mechanics is termed the `black hole information paradox' because the information in the initial wavefunction disappears from the the final configuration of the system. 

One may think that this problem is arising because we do not have a correct theory of quantum gravity. If that was the  case there would be no {\it paradox}; merely a observation that the details of planck scale physics need to be worked out better. 
But Hawking's paradox is so important because it appears to require no details of quantum gravity at all. The only assumptions that go in are (i)  usual laboratory physics holds in gently curved spacetime and (ii) the black hole  horizon is a regular region with gentle curvature. 

Given Hawking's result, one may wonder why physicists have not been {\it more} worried about the paradox. The situation is actually even more confusing. Many relativists are convinced that evaporation of black holes leads to information loss or `remnants' (planck sized objects that `lock up' information). String theorists, on the other hand believe in neither of these outcomes; they expect that the black holes evaporation is like the burning up of a piece of paper; the initial wavefunction of the hole {\it does} become an ordinary final state wavefunction of the emitted quanta, so there is no `information loss'. 

If the string theorists do not see a problem, then what is their answer to Hawking's argument? It turned out that string theorists quite generally believed that Hawking's argument was {\it faulty}: Hawking only did a leading order computation showing information loss, and if subleading terms were carefully computed, the problem would go away. But if subleading corrections were indeed important, why did relativists continue to believe in information loss?

In 2005 Hawking himself conceded a long held bet with Preskill, arguing that small corrections from subleading saddle points in the Euclidean path-integral could remove the problem \cite{hawkingreverse}. But Hawking's co-signer on the bet, Thorne, refused to concede; after all Hawking had not shown any clear way in which small corrections {\it could } remove the difficulty. 

So who is right?

In 2009 an inequality was derived which proves that small corrections {\it cannot} resolve the problem \cite{cern}. What then is the resolution? It turns out that what breaks down is assumption (ii) which stated that black holes  have regular horizons. This assumption arose from  many years of  efforts by relativists to find deformations of the black hole geometry. The failure of these efforts led to a belief: `black holes have no hair'; that is, the only state for the black hole appeared to be the standard one with a round, gently curved horizon. We will see that in string theory we can in fact find `hair'; it turns out that there are {\it no} regular horizons for energy eigenstates in string theory. The `hair' arises from a nonperturbative twistings of the compact directions over the noncompact ones. For the simplest holes we can count the all such states and find that their number agrees with the Bekenstein entropy.

The construction of these states (called `fuzzballs') resolves the information paradox, since they invalidate Hawking's premise of a smooth horizon.  Further, we can explicitly compute the emission from the simplest fuzzballs and see the information of the microstate being imprinted on the outgoing radiation, though the coarse-grained radiation {\it rate} agrees with the Hawking radiation rate expected for that state. 

But we can still wonder if any aspect of our original classical intuition still holds; i.e.,  is there a sense in which the traditional black hole geometry is still valid? We will see that the key to understanding this issue is to note a separation of energy scales.  For energies of order the temperature $T$ of the hole, the dynamics at the horizon sees the detailed difference between microstates; this is of course required if the radiation is to carry away information. But experiments at $E\gg kT$ excite collective modes of the fuzzball, which are insensitive to the precise state of the fuzzball. Further, correlation functions derived from such collective modes   agree to leading order with correlators computed in the traditional black hole geometry.

In this article we will give an overview of these developments, and end with a conjecture on the implications of this physics for early Universe Cosmology.

\section{The entropy problem}

Consider a black hole of mass $M$. Take a box of gas, containing an entropy $S_{matter}$, and throw it into the hole. The gas disappears from view, taking its entropy with it. Have we reduced the entropy of the universe and thus violated the second law of thermodynamics?

The general answer to this would be {\it no}, it should not be so easy to violate the second law. If we throw the box of gas in a trash can, we do not decrease overall entropy; the entropy of the can goes {\it up}. Of course we cannot easily look inside a black hole to check if the entropy is there, but somehow we would still like to believe that when the black hole swallowed the box of gas, its entropy increased. The mass of the hole certainly  goes up, and so does its radius $R={2GM\over c^{2}}$. Suppose we associate an entropy with the black hole equal to \cite{bek}\cite{hawking}
\be
S_{{bek}}={c^3A\over 4\hbar G}={A\over 4G}
\label{one}
\ee
where $A$ is the surface area of the hole and in the second step we have set $c=\hbar=1$. Then detailed computations show that
\be
{dS_{{bek}}\over dt}+{d S_{matter}\over dt}\ge 0
\ee
for all physical processes involving the black hole. This suggests that the `Bekenstein entropy' (\ref{one}) is indeed the entropy of the black hole.

But from the work of Boltzmann we have learnt that, for usual systems, the entropy equals $\log{\cal N}$, where ${\cal N}$ is the number of microstates for a given total energy. This suggests that the black hole should have $Exp[S_{{bek}}] $ microstates. This is where we encounter our first puzzle. People tried to look for these microtstates by looking for small distortions of the usual Schwarzschild geometry
\be
ds^2=-(1-{2M\over r})dt^{2}+{dr^2\over 1-{2GM\over r}}+r^{2}{d\Omega_2^{2}}
\label{two}
\ee
One could then hope that these approximately similar objects would give the $Exp[{\cal N}]$ microstates needed to account for the entropy $S_{bek}$.
{\it But people could find no such deformations} \cite{nohair}. If we take any small perturbation of 
(\ref{two}) then the energy in the perturbation either flows off to infinity or falls to the center of the hole, and the horizon settles down to its standard round shape as before. Repeated failure to find deformations of (\ref{two}) came to be called  a `theorem': {\it black holes have no hair} -- they have only the standard `bald' shape. 

If a black hole of mass $M$ has only one state, then its entropy would be $\ln 1=0$. This is in sharp contrast to the expression (\ref{one}), which implies $10^{10^{77}}$ states for a solar mass black hole.

This is a puzzle, but not really  {\it problem} for physics.  It could just be that the differences between states all lie at the central singularity $r=0$, where they are buried in planck scale physics. Or perhaps entropy is not given by a count of states in gravitational systems. But the next problem encountered with black holes was very serious, and could not be explained away by our ignorance of quantum gravity.

\section{The information paradox}

\begin{figure}[htbp]
\begin{center}
\includegraphics[scale=.55]{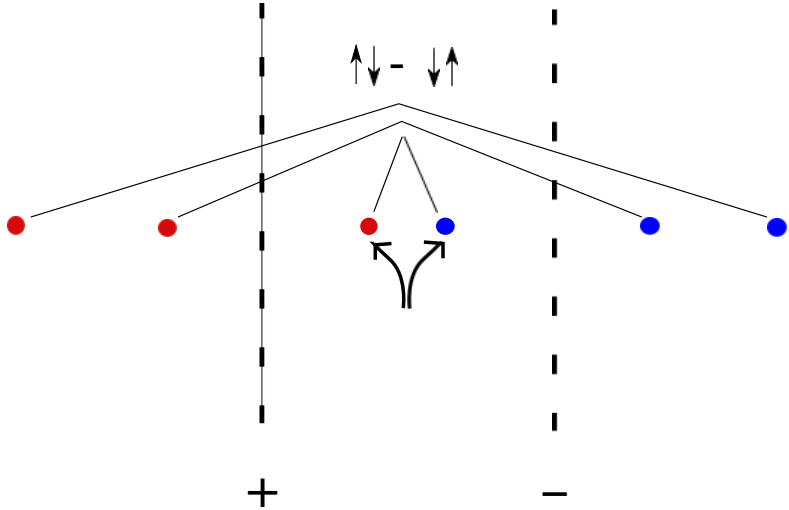}
\caption{{Electron positron pairs are created from the vacuum, and pass through the positive and negative grids. The two members of each pair are entangled with each other, generating an entanglement entropy $S_{ent}=N\ln 2$ between the left and right sides of the figure.}}
\label{f1}
\end{center}
\end{figure}

To understand the paradox found by Hawking, it is useful to first look at the {\it Schwinger process} in electrodynamics. There is no paradox with the Schwinger process, but this process is so similar to particle production in black holes that it is useful as a warm up model. Thus consider two parallel wire mesh plates, one carrying positive charge and one negative (fig.\ref{f1}). There is an electric field $\vec E$ between the plates. The normal vacuum always contains vacuum fluctuations -- pairs of electrons and positrons are created  and quickly annihilated. But in the presence of the electric field, a pair can be created, and before it has a chance to annihilate, get {\it separated} -- the electron getting pulled to the positive mesh and the positron to the negative mesh. These particle pass through the respective meshes and fly apart,  and we have obtained a particle pair  from the vacuum. The process keeps repeating, as long as the field $\vec E$ is maintained.

The important aspect for us is the state of a created pair. Let us assume that the pair is created in a spin singlet state:
\be
|\psi\rangle_{{pair}}={1\over \sqrt{2}}\Big (|\uparrow\rangle_{e^{-}}|\downarrow\rangle_{e^{+}}-|\downarrow\rangle_{e^{-}}|\uparrow\rangle_{e^{+}}\Big )
\ee
Then the electron and positron are in an {\it entangled} state, with an entanglement entropy $S_{{ent}}=\ln 2$. After $N$ steps of pair creation, the state to the left side is entangled with the state on the right side with $S_{ent}=N\ln 2$. 

This computation is straightforward and does not lead to any  problem: entangled states are common in physics, and all we have seen is that the Schwinger pair creation process naturally leads to entangled states. But now we will see how a similar process occurs around black holes (where the gravitational field replaces the electric field above) and this time there {\it will} be a problem.

\begin{figure}[htbp]
\begin{center}
\includegraphics[scale=.55]{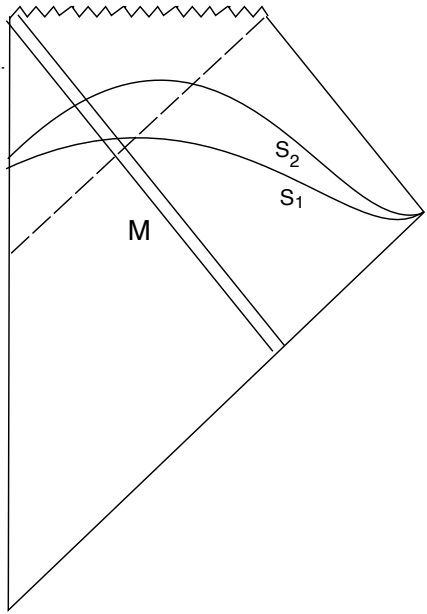}
\caption{{The Penrose diagram of a black hole formed by collapse of the `infalling matter'. The spacelike slices satisfy all the niceness conditions N.}}
\label{fthree}
\end{center}
\end{figure}

\begin{figure}[htbp]
\begin{center}
\includegraphics[scale=.18]{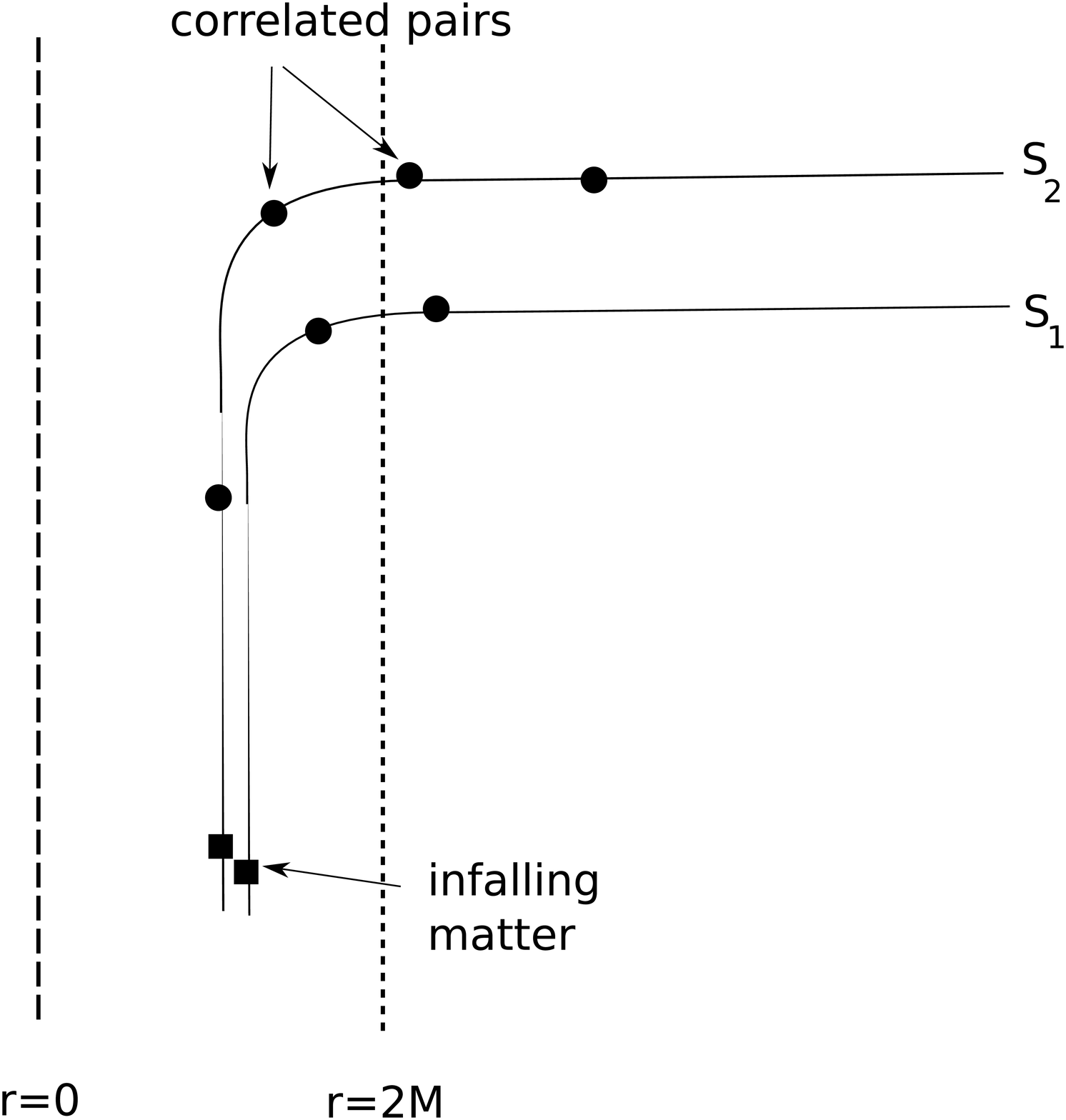}
\caption{{A schematic set of coordinates for the Schwarzschild hole. Spacelike slices are $t=const$ outside the horizon and $r=const$ inside. Assuming a solar mass hole,  the infalling matter is $\sim 10^{77}$ light years from the place where pairs are created, when we measure distances along the slice. Curvature length scale is $\sim 3 ~km$ all over the region of evolution covered by the slices $S_i$.}}
\label{ftwo}
\end{center}
\end{figure}

The crucial property of the black hole geometry is the existence of a horizon at $r=M$,  where the coefficient of $dt^{2}$ and $dr^{2}$ change sign. To study evolution, we need to foliate spacetime by {\it spacelike} slices (figs.\ref{fthree},\ref{ftwo}). For $r>3M$, say, we can use the normal spacelike slice $t=const$. Inside the hole, however, $t=const$ is timelike, and a spacelike slice would be $r=constant$. Let us take $r=M$, which is not too near near the singularity $r=0$ and not too near the horizon $r=2M$. The parameter $t$ measures distances along this spacelike alice, and we see that there can be {\it an infinite amount of proper length on the spatial slice inside the hole}. This is the feature that distinguishes the black hole from, say, a neutron star. We can connect the $t=const$ part of the slice with the $r=const$ part using a smooth `connecter' segment -- this can be seen from the Kruskal coordinates covering the black hole geometry.

Let us see the consequences of this curious foliation. To study evolution we have to move from one spacelike slice to a later one. Outside the hole, we make the new slice by taking $t\r t+\Delta$. For the part $r=const$, let us not evolve al all; evolving forward in time means evolving closer to $r=0$, where we do not want to go. We keep unchanged the intrinsic geometry of the connector slice. To join the connector to the $r=M$ part of the slice, we see that we have to take  a {\it longer} segment of the $r=M$ slice. Thus the later slice in the foliation is obtained by `stretching' part of the earlier slice. The black hole geometry had appeared time-independent in the Schwarzschild coordinates (\ref{two}), but those coordinates failed at the horizon. Any complete foliation by Cauchy slices of the black hole geometry {\it will necessarily be time-dependent}. This time dependence of the geometry prevents the system from settling down to a given vacuum state: since the vacuum is different on each slice, the vacuum state on one slice becomes the vacuum on the next slice {\it plus} some particle pairs. 

Thus we get  pair creation similar to pair creation in the Schwinger process. One particle is created on the part of the slice outside the horizon, and one inside. Each time we evolve the outer part of the slice by $\Delta t\sim M$, the middle of the slice stretches; the earlier created pairs get pushed further apart, and a new pair gets pulled out of the vacuum in the middle of the slice. The crucial point again is the state of the created pair: it is an entangled state. This entanglement may come from the spin of the created quanta, or from an internal degree of freedom like isospin, but even for particles carrying no quantum numbers we get an entanglement. At each step of `stretching' there is an amplitude to create no pairs, an amplitude to create one pair, an amplitude to create two pairs, etc. Restricting to the first two possibilities, we write the state schematically as
\be
|\psi\rangle_{pair}={1\over \sqrt{2}}\Big ( |0\rangle_{n}|0\rangle_{n'}+|1\rangle_{n}|1\rangle_{{n'}}\Big )
\label{psipair}
\ee
where $n, n'$ refer to quanta on the inner and outer parts of the slice at the $n$th step in the evolution. As in the Schwinger process, with the state (\ref{psipair}) we get an entanglement entropy $S_{ent}=N\ln 2$ after $N$ steps of evolution. (See \cite{giddingsnelson}) for a detailed computation.)

The important issue comes when we consider the endpoint of the pair creation process.   In the   Schwinger process, 
if we had taken two solid plates instead of wire meshes and let $\vec E$ be created by the charge on these plates, then the created pairs would slowly deplete the charge on the plates. When $\vec E$ goes to zero, the process stops, and we are left with two heavily entangled plates. In the black hole, the outer members of the pair float off to infinity, carrying energy. The inner members of the pair have {\it negative} energy, so they reduce the mass of the hole. What is the endpoint of this process? There are two possibilities:

\b

(i) The evaporation stops when we reach a planck sized remnant. The outer and inner spaces are entangled with a large entanglement entropy $S_{{ent}}
=N\ln 2$. This means that the remnant must have at least $2^{N}$ internal states. But $N$ can be made arbitrarily large, since we could have started from an arbitrarily large hole. Thus we are forced to admit an unbounded number of states in a finite energy range (planck mass) in a finite volume (planck length). It seems hard to imagine a field theory that could accommodate this requirement. 

\b

(ii) The hole evaporates away completely. In this case the outer members of the pairs are entangled, but there is nothing left that they are entangled {\it with}. These quanta can therefore not be described by any wavefunction; only by a density matrix. Thus an initial pure state (the gas making up the collapsing star) evolves to a black hole, and after evaporation, into something that can only be described by a mixed state. This outcome, initially proposed by Hawking, destroys the underlying structure of quantum mechanics where wavefunctions evolve to wavefunctions by a unitary evolution process.

\b

Why did we run into a problem with the black hole, and not with the Schwinger process? In both cases we create an entangled state, but in the black hole case we deplete away to zero the energy of one of the two entangled sides. If we have no energy, it is difficult to imagine having any states, and that is the source of the problem. But how could the black hole mass go to zero? The hole contained all the stuff that went in to make the hole in the first place, as well as all the infalling members of the created pairs. But the latter had negative energy, and this is how the total mass of the hole can go to zero. How can some quanta have negative energy? This is where the special property of gravity comes in: the gravitational potential is attractive, so an object of mass $m$ placed near a hole of mass $M$ has, in a Newtonian estimate, an energy $E=mc^{2}-{GMm\over r}$. For $r \lesssim {GM\over c^{2}}$ we get $E<0$; a proper relativistic computation supplies a factor of $2$ giving the critical distance as $r={2GM\over c^{2}}$. Thus quanta inside the horizon can have negative energy, and this negative sign can be traced back to the attractive nature of gravity. 

Before moving on, let us note that the entropy puzzle and the information problem are closely connected. Normal bodies do not have their mass all sitting at $r=0$; their structure is spread over their volume,  so we can explicitly see their microstates and count them. When it comes to radiating energy, microstates radiate from their surface, where the details of the microstate gets encoded in the emanating radiation. In the black hole, we have a vacuum solution outside $r=0$. Thus we cannot count microstates, and radiation is emitted by pair production from the vacuum; this creates the entangled state that leads to the information paradox.

\section{The issue of small corrections}

Relativists tried hard to bypass the above paradox, but could find no way around it. Finding `hair' at the horizon would have helped: if the geometry in the horizon region was not the vacuum geometry, then perhaps the pair creation would be altered in some way that would not lead to a monotonically rising entanglement between the inside and the outside of the hole. But the `no-hair theorem' remained firm. Reluctantly, many of them became believers in either the existence of remnants or in the inevitability of information loss.

The string theory community, however, were not overly concerned about  the problem. String theory seemed to be a good quantum theory of gravity at low energies, so it was hoped that it would take care of any problems in more complex situations, like black holes. But what about Hawking's explicit computation leading to remnants/information loss?

Most string theorists believed that Hawking's argument was {\it wrong}. There were a few different points of suspicion. As it turned out, these suspicions were unfounded and the original Hawking argument was correct, but let us take a moment to review these arguments because they confused the issue for a while.

\b

\subsection{Transplankian modes}

In the way Hawking's original computation was presented, it appeared that later and later pairs would be produced from vacuum modes that started off with shorter and shorter wavelengths. Since these initial wavelengths would be shorter than planck length, and we do not really know physics below the planck length, it was believed that Hawking's computation relied on an oversimplified model of the quantum vacuum at transplankian energies.

This fear is, however, unfounded, since one can redo Hawking's argument in a way that does not involve transplanckian physics at all. All we have to do is use a `good slicing' at the horizon, like the one we used in our above explanation of Hawing radiation. Let the black hole be solar mass, with a radius $\sim  3~ km$. We follow the evolution of a fourier mode straddling the horizon from the time it is wavelength 1 fermi to the time its wavelength is $\sim 3 ~km$: a range where we know quantum physics very well. Now we ask the following. When the wavelength was 1 fermi, was the mode in the vacuum state or an excited state? If it was excited, we would have nuclear density at the horizon, not the vacuum, and we have to explain how we got such `hair' in place of the metric (\ref{two}). If the mode was in the vacuum state, then its evolution on the gently curved metric (\ref{two}) is {\it known}, and so we have no choice but to get the entangled pair when the wavelength becomes $\sim 3$ Km. Thus if we are given the `no-hair theorem', we can arrive at Hawking's problematic entanglement without any mention of transplanckian physics.

\subsection{Small corrections}

A  second   confusion arose from the notion of small corrections. The argument goes as follows. Hawking did a leading order computation of pair creation, which gave a large entanglement $S_{ent}$ between the outer and inner quanta. But at each creation step there will be {\it some} small corrections do to quantum gravity effects. These corrections have to be small, since we know the evolution of low energy modes at the smooth horizon to a  very good approximation. But the number of created pairs is very large, of order $\Big ({M\over m_{pl}}\Big)^{2}$. Small corrections to each pair, cumulated over a large number of pairs, may lead to a final state where $S_{{ent}}\approx 0$, which would evade the problem.

If small corrections {\it could} indeed reduce $S_{ent}$ to zero, there would be no Hawing paradox; everyone  would readily admit the existence of subleading corrections to Hawking's entangled pair state. The problem was that no one had ever shown an example where  small corrections did lead to a nonentangled state. And string theorists did not explicitly produce such an example; they implicitly believed that {\it something} should make 
Hawing's problem go away, and since the black hole geometry (\ref{two}) looked inevitable, it must be the small corrections from quantum gravity effects that were taking care of the problem. 

It was such an argument of small corrections that led Hawking to reverse his view in 2005 and agree that information would probably not be lost \cite{hawkingreverse}. He considered the Euclidean path integral,and noted that there would be a leading saddle point giving semiclassical physics, and subleading saddle points giving small corrections. (In fact such corrections are exponentially small in the mass of the hole.) It was then argued that taking these subleading effects would remove the original entanglement problem. 

But most relativists could see no power in this argument: if the subleading saddle point could change things in the Euclidean language, then what is the analog of this change in the Lorentzian section where the problem is actually posed? We have not changed the leading order semiclassical evolution, so why is it clear that the large entanglement will go away? As mentioned in the introduction, Hawking surrendered his bet to Preskill, but Thorne, Hawking's co-signer, did not.

As it turned out, the small corrections argument is {\it wrong}. Suppose that at each step of evolution the state of the entangled pair can differ from Hawking's leading order pair state only by a small admixture of the other state making up the Hilbert space of the pair 
\be
|\psi\rangle_{{pair}}= \alpha\Big ( |0\rangle_{n}|0\rangle_{n'}+|1\rangle_{n}|1\rangle_{{n'}}\Big )+\beta \Big ( |0\rangle_{n}|0\rangle_{n'}-|1\rangle_{n}|1\rangle_{{n'}}\Big )
\ee
with 
\be
|\beta|<\epsilon, ~~~\epsilon\ll 1
\ee
Then the total entanglement entropy can have only a small change from Hawking's leading order result \cite{cern}
\be
{\delta S_{ent}\over S_{ent}}<2\epsilon
\label{theorem}
\ee
The proof of this result involved using strong-subadditivity of quantum entanglement entropy; a property that is somewhat nontrivial to prove.  But something like this inequality was intuitively being assumed by the relativists who felt that Hawking's argument could not be bypassed by small corrections. The important thing to note about (\ref{theorem}) is that tis inequality  involved the size of the correction  $\epsilon$, but not the dimension $d=2^{N}$ of the Hilbert space of states. While there are a large number of entropy inequalities for quantum entanglement entropy, most do involve $d$; here we had to find one that does not, since $d$ is very large in our problem and any combination of $\epsilon$ and $d$ cannot be assumed small.

 \subsection{AdS/CFT duality}
 
 A widespread belief in the string community was that the paradox has been 
 bypassed in string theory because of the discovery of AdS/CFT duality \cite{maldacena}. It had been found that correlators in anti-de-sitter gravity backgrounds could be reproduced by correlators in a field theory {\it without} gravity. Since the latter is a manifestly unitary theory, it might seem that there cannot be any information loss in gravity either. But in that case, what is the error in the Hawking computation? It turned out that the agreement of correlators in the duality is typically checked for {\it low} energy processes, where black holes do {\it not} form. For high energies, the string theorists again wrote a black hole metric analogous to (\ref{two}), and studied only its thermodynamic properties like entropy. The relativists can (and did) complain that this solves nothing: there has never been any problem with unitarity for processes where black holes do {\it not} form, and for the case where they do, the string theorists have shown no way to solve Hawking's problem. Should we not just say that the AdS/CFT map fails when black holes form?  
 
 It turned out that the string theorists answer to this criticism was again based on the hope that small corrections to Hawking's leading order computation would remove the paradox altogether. But the inequality (\ref{theorem}) shows that this hope was false. The source of confusion can be traced to a result of Page \cite{page}. Page's observation is, in itself, completely correct. He notes that when a piece of paper burns away to photons, the information of the paper is certainly contained in the photons, but almost no information can be obtained unless one examines at least half the emitted photons. The confusion arose when this observation was used to argue as follows: `if the information in the paper can be `delicately encoded' in the emitted photons, then cannot small corrections to Hawking's computation `delicately encode' the information of the hole?' This argument is false, and the error be located once we try to make precise what `delicately encoded' means.  For the burning paper, the photon emitted at each step has a spin that depends on the spin of the atom emitting it; thus there is a  2-dimensional space of possibilities to consider at each step, with the choice of state determined by the data in atoms of the paper.  In black hole evaporation the state at each emission step is close to the same state (\ref{psipair}) at each step (independent of the data in the hole). Thus we are restricted to a 1-dimensional space  at leading order, and small corrections cannot help much, as established by (\ref{theorem}).
 
 \section{Fuzzballs}
 
 Let us summarize what we have seen so far. Suppose we wish to believe that the correct theory of quantum gravity has no information loss or remnants. Then there are two results which, taken together,  stand in our way:
   
 \b
 
 (i) The Hawking argument, which relies only on the existence of  the black hole metric (\ref{two}). The result (\ref{theorem}) proves that the argument is unaffected by any kind of small correction to the leading order computation.
 
 \b

 (ii) The `no hair theorem', which summarizes the fact that many efforts to find deformations of the black hole solution (\ref{two}) did not succeed. 
 
 \b
 
 As we will see below, it turns out that in string theory (ii) will fail: we {\it do} find hair for black holes in string theory. To understand the construction, let us recall some principal features of the theory. Strings are consistent only in 10 dimensions; thus to get lower dimensional physics we must compactify some directions. The fundamental quanta of the theory (and their interactions) are completely fixed; there are no free parameters in the theory. Nothing can be added or deleted from the theory without causing an inconsistency. 
 
 Thus to make a black hole we must use the objects already in the theory. Consider a compact circle of radius $R$. We can wrap a string $n_w$ times around this circle; from the point of view of the noncompact directions this gives an object with mass and `winding charge',  and for large $n_w$ we might hope to get a black hole . It turns out that the tension of the wrapped string causes the compact direction to `pinch', and the horizon area becomes zero.  The problem can be remedied by adding $n_p$ units of momentum to the string; the energy of a momentum mode scales as $1/R$, so the momentum pushes the circle radius towards larger values. Between the tension of the winding modes and the pressure of the momentum modes,  the compact circle settles to a finite nonzero radius, and we can hope to make a black hole solution.  
 
 Now we observe an interesting  fact: a string carrying momentum has a large {\it entropy}. The momentum on the string is carried by {\it travelling waves} on the string, and we can partition the total momentum among different harmonics in different ways.  The entropy of this microscopic count of states turns out to be \cite{sen}
 \be
 S_{micro}=4\pi\sqrt{n_wn_p}
 \ee
On the other hand we can make a gravitational solution with `winding charge' $n_w$ and `momentum charge' $n_p$ and compute the entropy from its horizon area. In this case stringy corrections to the Einstein actions must be included, and the correct analog of (\ref{one}) is the `Bekenstein-Wald' entropy. One finds \cite{dabholkar}
\be
S_{bek-wald}=4\pi\sqrt{n_wn_p}=S_{micro}
\label{ten}
\ee
This system is called the 2-charge hole. A similar computation can be carried out by adding more kinds of charges, and similar results are obtained; in particular the celebrated Strominger-Vafa computation for the 3-charge hole was the first case of a result like   (\ref{ten}) with full agreement of numerical factors \cite{sv}.
 
 All this does not, however help us with the information problem: the black hole geometry used to compute $S_{bek-wald}$ is just the traditional spherically symmetric solution with a horizon and central singularity, so the metric is not fundamentally different from (\ref{two}). But let us look in more detail at the states we counted in $S_{micro}$. It is a basic fact of string theory that the string has no longitudinal waves; thus we need a transverse vibration profile to describe the momentum on the string. Let us open the multiwound string to its covering space, where it now has a length $2\pi n_w R$. The simplest vibration profile is to let the string describe one turn of a uniform helix in this covering space. In our actual space, the string  looks like a `slinky' bent into the shape of a circle. The metric of this string profile can be computed; it is axially symmetric, with no horizon. We can use the S,T dualities of string theory to map the winding and momentum charges to D1 and D5 brane charges; in that case the metric turns out to have no singularities whatsoever \cite{lm3}. 
 
 \begin{figure}[htbt]
\hskip .5 in\includegraphics[scale=.25]{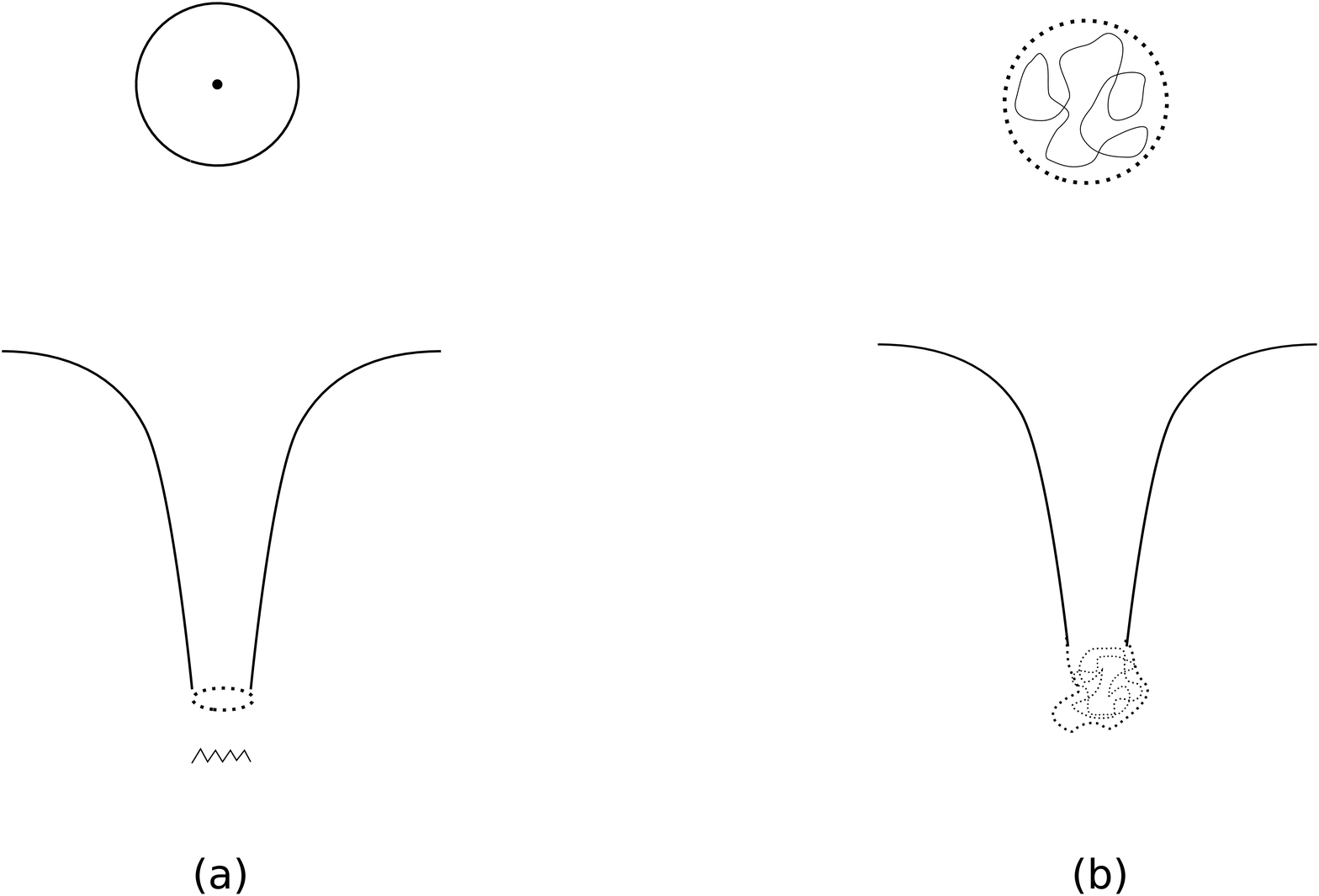}
%
%
\caption{(a) If the string winding and momentum excitations could sit at a point, then we would get the usual black hole; in the lower diagram the geometry is shown with flat space at infinity, then a `throat', ending in a horizon with a singularity inside. (b) The string cannot carry the momentum without transverse vibrations, and thus spreads over a horizon sized transverse area. The geometry depicted in the lower diagram has no horizon; instead the throat ends in a `fuzzball'.}
\label{matftfive}       
\end{figure}

 Should we be surprised that one particular state of the 2-charge system turned out to have no horizon at all? The metric of this solution can be found from known families of solutions for rotating black holes: when one increases the rotation to a critical point, the horizon disappears, and for more rotation one gets a naked singularity. The special solution noted above corresponds to the critical rotation, and gives the maximally rotating state of the 2-charge system \cite{bal,mm}.  It seems that we have simply pushed the problem to the remaining solutions which have lower rotation; the axially symmetric ansatz would generate a horizon (after $R^2$ terms in the action are used) for these lower rotation solutions.
 
 Since we know that all our solutions are generated by a vibrating string, let us look at the other profiles of vibration for the string, and see what metric they generate. It turns out  that {\it no} vibration profile generates a horizon; what happens instead is that in general {\it we lose axial symmetry} \cite{lm4}. If fact it is easy to consider vibration profiles for the string which have {\it zero} total angular momentum. The corresponding gravitational solutions are not spherically symmetric: it is impossible to make put a transverse wave on a string while maintaining spherical symmetry. To summarize, the situation for the 2-charge extremal hole is as follows:
 
 \b
 
 (i) If we take a spherically symmetric ansatz for the metric, we get a solution with horizon and a central singularity; the horizon area gives the entropy of the system \cite{dabholkar}. 
 \b

 (ii) The actual solutions corresponding to the entropy can all be constructed, and in general they have no spherical or axial symmetry, and none of the solutions have a horizon.  In particular the solution in (i) does not correspond to any of the microstates of the \cite{lm4}.
 
 \b
 
 (iii) The space of these gravitational `fuzzball' solutions can be quantized and shown to yield the entropy of the system \cite{rychkov}.
 
 \b
 
 (iv) The size of the typical `fuzzball' solution is such that its boundary area satisfies ${A\over G}\sim S_{micro}$ \cite{lm5}. 
 
 \b
 
 This completely solves the information problem for this simple 2-charge hole, which now has its internal information manifested in its detailed structure.

  Why had we not found such non-spherically symmetric solutions before? It turns out that while people had looked for {\it perturbative} deformations around the spherically symmetric hole, the above mentioned solutions are {\it nonperturbative} constructions involving the compact directions. A compact circle can be nontrivially fibered over the noncompact directions to make a `Kaluza-Klein monopole'. The fuzzball solutions are generalizations of such fibrations; there is no net `monopole charge', but the phase space of `monopole-antimonopole solutions' is vast, and gives the microstates of the hole.
  
  This fuzzball construction has been extended to cover large family of microstates of more complicated holes carrying 3 or 4 charges \cite{fuzzball2}. While most solutions constructed are extremal, some families of non-extremal solutions have been constructed as well. Remarkably, these nonextremal solutions possess unstable `ergoregions' in their interior \cite{ross, myers}, which leak radiation at a rate that is exactly the rate of Hawking radiation expected for those microstates \cite{cm1}. The radiation is not `information-free' -- its spectrum is imprinted with the details of the radiating fuzzball. The microstate here is very simple, and so not a very generic one for the black hole; nevertheless what we have here is  an  explicit  example of a black hole microstate radiating its information through its Hawking radiation.

  \section{Gravitational collapse}

  What we have seen is that energy eigenstates in string theory do not have a traditional horizon. But in classical general relativity we can take a shell of mass $M$, and follow its gravitatonal collapse. The shell appears to pass smoothly through the radius $r=2M$, creating a horizon behind it. How can fuzzballs modify this process, which seems to be so very classical?

 To answer this, imagine the process where the collapsing shell {\it tunnels} into one of the fuzzball states \cite{tunnel}. The amplitude for such a process will be tiny, given by $Exp[-S_{cl}]$, where $S_{cl}$ is the Einstein action ${1\over 16\pi G}\int R \sqrt{-g}d^4 x$. To estimate the order of $S_{cl}$, we set all length scales to be $\sim GM$. Then $R\sim (GM)^{-2}$, and $\int \sqrt{-g} d^4 x\sim (GM)^4$, so we get
 \be
 S_{cl}\sim {1\over G } (GM)^{-2}(GM)^4\sim GM^2\sim \Big ({M\over m_{pl}}\Big )^2
 \ee
 Thus the tunneling amplitude is indeed very small, as expected for any tunneling process between two macroscopic configurations. But there are $Exp[S_{bek}]$ fuzzball states that we can tunnel to, so we must multiply the tunneling probability by the number of states ${\cal N}$, where
 \be
 {\cal N}=Exp[S_{bek}] = Exp[{A\over 4G}]=Exp[4\pi GM^2]\sim Exp[\Big ({M\over m_{pl}}\Big )^2]
 \ee
 We see that the smallness of the tunneling amplitude can be compensated by the largeness of ${\cal N}$. The tunneling time was estimated in \cite{rate}, and was found to be much shorter than the Hawking evaporation time. Thus we find that the wavefunction of the collapsing shell will change to a linear combination of fuzzball wavefunctions, and the radiation process will then process from the surface of these fuzzballs just like radiation from any normal warm body.

Interestingly, the quantum theory of black holes began with the  observation that black holes have much more entropy than normal bodies. In simple approaches like canonically quantized gravity, we do not see the degrees of freedom corresponding to this entropy, and the traditional horizon leads to information loss/remnants. But in string theory we do see the different microstates as explicit horizon-sized wavefunctions. We then find that the wavefunction of any initial configuration (e.g. a collapsing shell)   spreads quickly over this vast  phase space, and classical evolution becomes invalidated. Put another way, any path integral amplitude has a classical action in the exponential, and a measure term which is typically smaller by a power of $\hbar$. For the black hole this measure factor (which gives the phase space volume) is so large that it competes with the classical action term, making the black hole interior a quantum object.

What is the special property of string theory that allowed us to have these macroscopic scale quantum effects? The key appears to a phenomenon called fractionation, which gave the first hint of a fuzzball structure for black hole microstates. Consider a string wrapped on a circle of radius $R$; we had used such a would string in our construction of 2-charge black holes. Momentum modes along this circle have energy given in units of ${1\over R}$. Now imagine that the string is multiwound, with winding umber $n_w$; thus it has a total length $n_w R$. The momentum excitations on the string now come in units of ${1\over n_w R}$; that is, in units that are ${1\over n_w}$ times the basic unit ${1\over R}$ \cite{dasmathur}. 

This may not appear very remarkable, since it is obvious that a longer string would have lower energy excitations. The interesting thing is that in string theory we have {\it dualities} that map the winding and momentum charge to other kinds of charges; for example we can map the winding and momentum  to D5 brane charge and D1 brane charge respectively. Then we will find that a D1 brane by itself has a very high tension (planck scale, if dimensionless couplings are taken order unity), but if the D1 brane is bound to $n_5$ D5 branes then its effective tension drops by a factor  ${1\over n_5}$ \cite{maldasuss}. 
Since low tension branes can stretch far,  we need to think carefully about the size of bound states containing a large number of strings and branes.

It had been traditionally believed that a bound state of string and branes would have a physical size of order planck length or string length. If this were the case, then for a large number of branes the horizon size would be larger than the physical size of the bound state, and we would have the traditional picture of a smooth horizon with a central singularity. But the phenomenon of fractionation alters the situation: as we increase the number of branes in the bound state we generate very low tension objects, and we should ask how far these objects can stretch. In \cite{emission} an estimate was made for the size of extremal bound state carrying D1, D5 and P charges. It was found that the size $D$ of the bound state was of order
\be
D\sim \Big ({g^2 \alpha'^3 \sqrt{n_1n_5n_p}\over RV}\Big )^{1\over 3}\sim R_s
\label{qwe}
\ee
where $g$ is the string coupling, $1/(2\pi\alpha')$ is the string tension, $n_1, n_5, n_p$ are the numbers of D1, D5, P charges, $R$ is the length of the compact circle and $V$ is the volume of the remaining four compact directions. The last step notes a remarkable fact: this estimated size $D$ is order the Schwarzschild radius of the D1D5P black hole! The fact that $D$ grows with the charges $n_1, n_5, n_p$ can be traced back to the phenomenon of fractionation.

Since the agreement in (\ref{qwe}) involves several different parameters, it appears to be more than a coincidence, and it motivated a search for concrete `fuzzball' constructions of black hole microstates. These constructions were first carried out for the simpler 2-charge D1D5 system, and now there has been considerable progress of the 3-charge and 4-charge systems as well. Overall, it appears quite clear that energy eigenstates in string theory alway have a size of order their horizon radius, and no traditional horizon forms at all.

\section{The infall problem}

Let us now turn to another common confusion with black holes. The  argument goes as follows: ``We know by the equivalence principle that nothing happens at the horizon. Then how can it be that in string theory we obtain nontrivial structure at the location of the horizon?''

Such an argument is misguided, though as we shall see, there {\it is} a useful question contained in it that we should answer. To see the error, note the following:  a body falling through the sky feels no force (and we can attribute this to the equivalence principle), but it lands with a sharp thud when it hits the earth's surface! The issue is of course clear: the equivalence principle holds only in empty space; when we reach the physical radius of an object (in this case, the earth) then other forces come into play and the equivalence principle tells us nothing.
For the black hole, {\it if} there is no structure at the horizon then we will fall smoothly through, but  if the natural `size' of the microstates is horizon scale  then there will be structure at the location of this horizon and there cannot be a {\it presumption} that `we see nothing at the horizon'. 

In string theory we have seen that there {\it is} structure at the horizon. We can still ask, however, if the traditional metric (\ref{two}) has any relevance to any process involving black holes. It turns out that there are two ways that (\ref{two}) is relevant:

\b

(i)  The Euclidean continuation of (\ref{two}) gives the `cigar' geometry (with no horizon) which is the saddle point of the path integral around black holes. This Euclidean saddle point can be used to compute the entropy of microstates, though each microstate in the Lorentzian section has no horizon and is not given by a metric like (\ref{two}). Such saddle point computations have been extended to yield precision counting of microstates in recent work with extremal holes \cite{sendabholkar}.

\b

(ii) We can ask: Does a freely falling observer `see' anything at the horizon? We will call this issue the `infall problem'. 
This infall problem is sometimes confused with the information paradox, but in fact the two problems are very different.  As we have seen above, the information paradox can only be resolved  if one finds structure at the horizon which will alter the evolution of Hawking pairs by order unity. Suppose that in a given theory of gravity we do find such structure. Then the question of what happens to an infalling observer is a  {\it secondary} issue: he may feel that he falls freely through the horizon, or he may feel a sharp obstruction, but in either case the worry is not one of {\it loss of unitarity}. 

Classical intuition suggests that there should be  some approximation in which an infalling observer sees the traditional geometry (\ref{two}). We 
will look for this approximation below, but let us note that this goal makes the information paradox and the infall problem, in a sense,  {\it opposite} problems. In solving the information paradox we seek to find ways in which different microstates behave {\it differently}, while obtaining traditional infall requires an approximation in which all generic microstates behave the {\it same}.

\b

In the rest of this section we discuss the infall problem and the approximation in which we obtain the traditional geometry (\ref{two}).\footnote{For other related approaches to this problem, see \cite{balasub}.}

\b

 A key point in this discussion will be the energy scales involved in the physics. Hawking radiation is composed of low energy quanta with $E\sim kT$, and we have seen that the evolution of such quanta is modified by order unity in a manner that depends on the details of the fuzzball structure at the horizon scale. But when we talk of observations made by an infalling observer we assume that he is looking at processes with $E\gg kT$, since for energies $E\lesssim kT$ the thermal bath of Hawking quanta would disrupt the validity of `vacuum physics' anyway. What we will argue now is that operators with $E\gg kT$ excite collective oscillations on fuzzballs that are insensitive to the exact choice of fuzzball microstate, and that the correlators of these collective excitations agree with correlators computed in the traditional geometry (\ref{two}). 

\begin{figure}[htbp]
\begin{center}
\includegraphics[scale=.65]{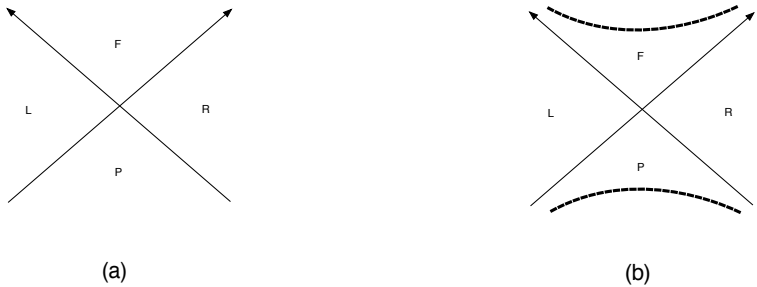}
\caption{{(a) Rindler space (b) The Penrose diagram of the extended Schwarzschild hole. The region near the intersection of horizons is similar in the two cases.}}
\label{fn5}
\end{center}
\end{figure}

To do this we will use ideas of Israel \cite{israel}, Maldacena \cite{eternal}, Van Raamsdonk \cite{raamsdonk} and others  and apply them in the context of fuzzballs \cite{plumberg}. We start with a simple observation about the behavior of correlators in Rindler space. As we can see from fig.\ref{fn5}, the central region of the extended Schwarzschild diagram is just like Minowski space, broken into the four `Rindler' quadrants. Thus we start by looking at  Minkowski space, and consider a scalar field on this space. The Minkowski vacuum state $|0\rangle_M$ can be written as a sum of contributions from states in the left and right Rindler wedges
\be
|0\rangle_M=C\sum_i e^{-{E_i\over 2}}|E_i\rangle_L|E_i\rangle_R, ~~~~~~~C=\Big (\sum_i e^{-E_i}\Big )^{-\h}
\label{split}
\ee
We now make a few observations. First, if we had an interacting scalar field, the states $|E_i\rangle_L, |E_i\rangle_R$ would be eigenstates of the {\it interacting} Hamiltonian. Second, an expansion like (\ref{split}) is expected for {\it any} field, and one field that is always present is nature is the graviton field $h_{ij}$. What is the analogue of (\ref{split}) for gravitational fluctuations $h_{ij}$? Note that the strength of gravitational interactions increases with energy.  The states $E_i$ have large local energy density  in the region close to the Rindler horizons, due to the large redshift near the horizons in the Rindler metric. Thus the states $|E_i\rangle_R$ for the gravitational field are expected  to be states with the following characteristics: (i) they should `live' in only the right Rindler quadrant (ii) they will have high local energy density near the Rindler horizons (iii) they will involve very nonlinear gravitational interactions near the horizons.

\begin{figure}[htbp]
\begin{center}
\hskip -1 in \includegraphics[scale=.60]{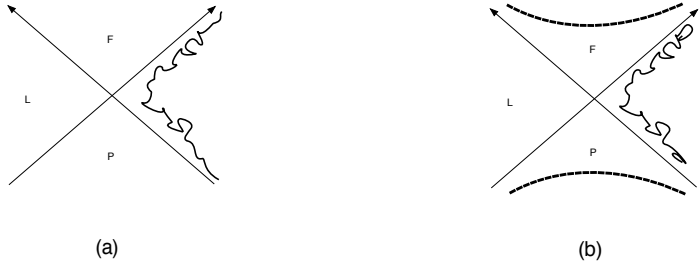}
 \caption{{Fuzzball structure for Rindler space (a) and the black hole (b). While the extended Schwarzscild metric has four regions separated by horizons, the fuzzball solutions of string theory occupy only the right (R) quadrant, and end in a quantum fuzz near the place where the horizon would have been (figure (b)). Carrying this notion over to Rindler space, we should think of the individual states of the R quadrant as ending in a quantum gravitational fuzz near the horizon (figure (a)).}}
\label{fn11}
\end{center}
\end{figure}

\begin{figure}[htbp]
\begin{center}
 \includegraphics[scale=.65]{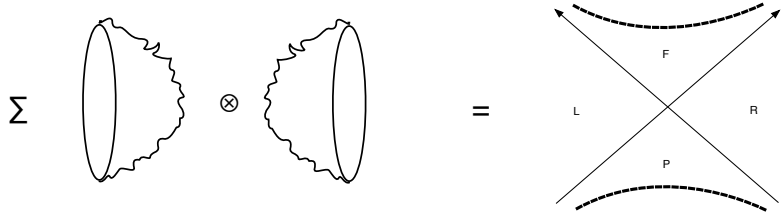}
\caption{{Black hole microstates are fuzzballs that `end' without forming a horizon. Summing over pairs of microstates (with appropriate weights) should give the geometry of the extended Schwarzschild hole.}}
\label{fn7}
\end{center}
\end{figure}

But these are just the characteristics of the fuzzball solutions that have been found! The fuzzballs end just outside the the place where the horizon would have occurred (fig.\ref{fn11}). Their structure at this location is over very short length scales and very nonperturbative: we have a complicated version of a  KK monople-antimonopole gas.  Thus it is natural to conjecture that the fuzzballs are just the solutions $|E_i\rangle_R$ appearing in the decomposition of the Minkowski vacuum for the gravitational field. We depict the relation (\ref{split}) for the graviton field pictorially in fig.\ref{fn7}.

Returning to (\ref{split}), consider the expectation value in the Minkowski vacuum of an operator $\hat O_R$ with support in the right wedge
\bea
{}_M\langle 0|\hat O_R|0\rangle_M&=&C^2\sum_{i,j}e^{-{E_i\over 2}}e^{-{E_j\over 2}}{}_L\langle E_i|E_j\rangle_L {}_R\langle E_i|\hat O_R|E_j\rangle_R\nn
&=&C^2\sum_i e^{-E_i}{}_R\langle E_i|\hat O_R|E_i\rangle_R
\label{qwe1}
\eea
Thus the expectation value in the Minkowski vacuum is given by  a thermal average over the Rindler states. 

Now consider the black hole. We had noted that the central part of the eternal Schwarzschild diagram (fig.\ref{fn5}(b)) was just like Minkowski space (fig.\ref{fn5}(a)). But when we make a black hole by collapse of a massive star, we do not get the eternal hole; the traditional expectation is that we get a Penrose diagram like that in fig.\ref{fthree}. We have argued that this latter diagram is misleading -- what we get instead is a fuzzball, which ends just outside the horizon region of the diagram. But we will get just {\it one} fuzzball  if we start with a star in a given pure state $|\psi\rangle$, not an ensemble of fuzzball states. Thus in the language used above, we get {\it one} of the states of the right Rindler wedge $ |E_k\rangle_R$, not an ensemble of all Rindler states.

If we compute the expectation value of $\hat O_R$ in this state we will get the analog of  ${}_R\langle E_k|\hat O_R|E_k\rangle_R$. This is not the same as the quantity ${}_M\langle 0|\hat O_R|0\rangle_M$ that we computed in (\ref{qwe1}); the latter is an ensemble average over right Rindler wedge states. But now we come to a basic fact of statistical mechanics:  for a {\it generic} state $|E_k\rangle_R$ and {\it appropriate} operators $\hat O_R$ we should be able to replace expectation values in the state $E_k$ by an ensemble average
\be
{}_R\langle E_k|\hat O_R|E_k\rangle_R\approx {1\over \sum_l e^{-E_l}}\sum_i e^{-E_i}{}_R\langle E_i|\hat O_R|E_i\rangle_R={}_M\langle 0|\hat O_R|0\rangle_M
\label{qwe2}
\ee
For example if we stick a thermometer in a beaker of water, then the rise of mercury can be computed using the actual state $|\psi_k\rangle$ of the water, {\it or} by using the ensemble average over such states; the result is expected to be the same to leading order. Here the state of water is assumed to be a {\it generic} state, and the operator measuring temperature is of the  `appropriate' type mentioned above. If on the other hand we had tried to measure the Brownian motion of one water molecule in the beaker, the result would depend very sensitively on which $|\psi_k\rangle$ we took, and we would {\it not} have the approximate equality (\ref{qwe2}).

In the black hole, it appears reasonable to assume that operators measured by infalling observers are of the `appropriate' type; i.e., they obey the approximation (\ref{qwe2}). This is because observers falling in from infinity impact the fuzzball hard (due to their large blueshift), and such high impact collisions `jump' the fuzzball state $|\psi_k\rangle$ into a whole band of neighbouring states. The physics then becomes similar to that involved in the `fermi golden rule': if an excitation is coupled to a large band of states then its absorption rate is insensitive to the precise state that the absorber started with; only the density of states and the average coupling are relevant. 

To summarize, correlation functions of appropriate operators in {\it one} Rindler state can be well approximated by an ensemble over all Rindler states (eq.(\ref{qwe2})). The relation (\ref{split}) tells us that the latter correlator will be the correlator computed in the Minkowski vacuum. Similarly,  correlators appropriate for infalling observers, measured in {\it one} fuzzball state can be approximated by the ensemble average over all fuzzballs. The similarity between Minkowski space and the central region of the extended Schwarzschild diagram (fig.\ref{fn5})) then tells us that such such an ensemble average  will be given by correlators computed in the extended Schwarzschild black hole. Since the extended Schwarzschild hole has no structure at the horizon, we have recovered a semblance of `classical physics' in the description of correlators of infalling observers. Note however that the low energy Hawking radiation quanta are like the atoms undergoing Brownian motion in our example of the beaker of water; their dynamics is {\it very} sensitive to the choice of fuzzball microstate $|\psi_k\rangle$. Thus we can have information  emerge in Hawking radiation quanta (which have $E\sim kT$) while we still have a role for a smooth horizon geometry for the high energy ($E\gg kT$) operators appropriate to infalling observers. 

Let us now make contact with the literature quoted above. The idea of using the two sides of the extended Schwarzschild diagram as the two copies of a system in real time thermal field theory goes back to Israel \cite{israel}. Using this notion, Maldacena \cite{eternal} noted that two copies of a CFT, in an entangled state, should be the dual of the eternal Schwarzschld hole. Van Raamsdonk \cite{raamsdonk} used these results to suggested a general idea that if we suitably entangle two copies of a {\it gravitational} system, then we get  a spacetime geometry that links the two systems. What we have done here is put these notions in the context of fuzzballs, and thus arrived at a picture where we can recover information in low energy Hawking radiation from the fuzzball structure while still having short time (crossing timescale) high energy ($E\gg kT$) physics  being reproduced by the traditional eternal black hole diagram 
with horizons.

\section{Implications for Cosmology: a conjecture}

Let us summarize what we have learnt.  The relativists who believed in information loss were correct about at least one thing: if there is no structure at the horizon, then we {\it cannot} get information out in Hawking radiation. This fact was however not universally accepted; many people believed  that small corrections to Hawking's calculation would alter his conclusion. The inequality (\ref{theorem}) proves this latter belief incorrect, and prevents one from claiming that the idea of AdS/CFT duality can by itself remove the paradox. 

One can then wonder if we {\it have} to accept information loss/remnants, since the `no hair theorem' appears to forbid any structure at the horizon. But the  `no hair theorem' was not really a `theorem'; it was just a statement that in traditional gravity theories one did not find `hair'. In string theory it turns out that there is indeed hair, arising from nonperturbative fibrations of the compact directions over the noncompact ones. The large phase space of these solutions invalidates classical dynamics on macroscopic scales, and the phase space measure of these `fuzzball' states causes a diffusion of the wavefunction away from its classical trajectory even for macroscopic collapsing shells.

Are there any consequences of this fact for other problems in physics? In the early Universe matter is crushed to high densities, much like in the classical black hole. Let us conjecture what might happen here using our experience from fuzzball physics.

\b

(A) If we take a gas of particles in a given volume $V$, the entropy scales with energy as $S\sim E^{d\over d+1}$ where $d$ is the number of space dimensions. Thus we have $S\sim E^\alpha$ with $\alpha<1$. But black holes have $S\sim E^\alpha$ with $\alpha>1$; for example in $d=3$ we have $S\sim E^2$. How does string theory reproduce such an entropy? Instead of using the energy $E$ to create a set of particles, we use $E$ to create a set of {\it extended} objects: strings, branes etc. For the black hole in $d=3$ space dimensions, we have four types of branes that can be wrapped on the six compact directions. Let there be $n_i$ branes of each type; thus $n_i\sim E$. The entropy of the configuration is given by $S\sim N_{int}^\h$, where $N_{int}$ is the number of {\it intersection points} of the different branes. We find   $N\sim n_1n_2n_3n_4$, which gives
\be
S\sim N_{int}^\h\sim (n_1n_2n_3n_4)^\h\sim E^2
\ee
reproducing the expectation for the 3+1 dimensional hole. Remarkably, this picture reproduces the entropy {\it with the correct numerical coefficient} for all the black holes constructed so far in string theory \cite{sv,hms}. For neutral holes (which have no net brane charge) we must still think in this fashion; we take branes and antibranes, vary the $n_i, \bar n_i$ till $S$ is maximized, and this maximal $S$ reproduces the Bekenstein entropy.

These computations suggest that at the high densities we must take the energy of the Universe and place it in a set of intersecting branes and antibranes: since this gives much more entropy than the traditional big-bang relativistic particle gas, we have vastly more states of this type than normal particle states. This idea was explored in \cite{cmfrac} where the intersecting brane configurations were termed the `fractional brane gas'. (Note that the `fractional brane gas' is completely different from the traditional `brane gas' \cite{bran} where we have essentially free (i.e. non-intersecting) branes with $S\sim E$.)

To summarize, let the Universe be a torus of volume $V$. We  take the energy $E$ in the Universe and use it to make states of the kind that give the entropy of black holes. At small enough $V$ (where the energy density is high), it appears reasonable that   the `fractional brane gas' states are the dominant type of state.

\b

(B) In \cite{cmfrac} it was assumed that once we have a fractional brane gas made of intersecting branes,  the entropy  is given by the  expression $S\sim N_{int}^\h$ which worked for black holes. But here we deviate from this assumption for the following reason. The above expression gave the entropy of black holes in empty space, where the states could expand to occupy a volume that optimized their entropy. As we know,  the black hole states swell to a size where their surface area satisfies the Bekenstein type relation ${A\over G}\sim S$.  The reason for this `swelling up can be seen explicitly for the simplest hole in string theory -- the 2-charge extremal hole discussed above, with states generated by a string carrying transverse vibrations. If we allow a very small transverse volume for these vibrations, there are very few states since we do not have enough phase space to carry too many orthogonal states. If we look at generic states (which account for the bulk of the entropy) then we find that they occupy a region whose surface area satisfies ${A\over G}\sim S$. There is no significant gain in entropy if we allow ourselves a larger  region than this, but what is interesting is what happens if we restrict ourselves to states which occupy a {\it smaller} region. In this case the entropy $S'$ of the allowed states  still satisfies a Bekenstein type relation
 $S'\sim{A'\over G}$, where $A'$ is now the surface area of the {\it smaller} region that we have allowed \cite{phase}. 
 
 In the case of the Universe, we do not have `empty space' to expand into; all the volume of the Universe is occupied more or less uniformly, and we have used this homogeneity to model our Universe as a torus of volume $V$. Suppose the Universe has energy $E$, which gives for the number of branes $n_i\sim E$. While the total possible entropy with these numbers is $S\sim \Big (\prod_i n_i\Big) ^\h$, all of the states corresponding to this $S$ may not be realizable in the given volume if $V$ is too small. Thus we can imagine that a small subset of states $\{\psi^{(0)}_i\}$ is allowed at volume $V^{(0)}$, a larger subset $\{\psi^{(1)}_i\}$ is allowed at a larger volume $V^{(1)}$, and so on. What is the consequence of this structure?
 
 \b
 
 (C) To guess what can happen, let us recall the evolution of a shell which was collapsing to make a black hole. In traditional gravity theories, this would generate a horizon which traps the shell, and we would get information loss/remnants. But in string theory the energy eigenstates are horizonless fuzzballs, and the wavefunction of the shell  spreads over the large phase space of these fuzzballs. Here it is important that the entropy of fuzzballs be large enough that the `tunneling into fuzzballs' competes with the classically expected  evolution \cite{tunnel,rate}. 
 
\begin{figure}[htbp]
\begin{center}
 \includegraphics[scale=.65]{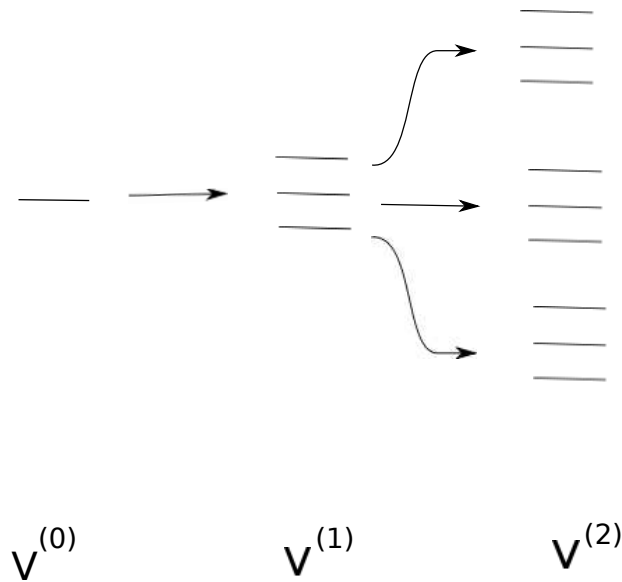}
\caption{The energy levels at different values of the Universe volume $V^{(k)}$. The states at any one level have a small amplitude $\epsilon$ per unit time to transition to the states in the neighboring volume. If the initial state is the one corresponding to  $V^{(0)}$, then there is a diffusion towards larger $V$ at a rate that depends on $\epsilon$ and the level density $\rho$.}
\label{fqone}
\end{center}
\end{figure}

 Let us put this picture of evolution with the picture we had for states in Cosmology in (B) above. We will make a toy model to illustrate the essential point of interest. In fig.\ref{fqone} we picture the states  $\{\psi^{(k)}_i\}$ for different volumes  $V^{(k)}$ of our toroidal Universe; we have taken the values  $V^{(k)}$ to be a discrete set for convenience. At the smallest volume $V^{(0)}$ we have taken just one state $\psi^{(0)}$. At the next larger volume we have a band of states $\psi^{(1)}_i$, with energy gap $\Delta E$. We allow a small amplitude $\epsilon$ per unit time for transition from $\psi^{(0)}$ to any of the $\psi^{(1)}_i$. Thus if the initial state of the Universe is the torus at volume $V^{(0)}$ with state $\psi^{(0)}$, then after some time $t$ this state can evolve to a linear combination of states $\psi^{(1)}_i$ at volume $V^{(1)}$. The rate of this evolution depends on the transition amplitude $\epsilon$, and the level density $\rho={1\over \Delta E}$. As in the black hole, we expect that $\epsilon$ is small (since we are dealing with a macroscopic system) and $\rho$ is large (since the density of `fractional brane gas' states is very high). The product of these two quantities became significant in the black hole case and drove the essential new physics, and we are looking for a similar effect here. 
 
 If we just had the states $\psi^{(0)}$ and $\psi^{(1)}_i$ then we could compute the evolution using the `fermi golden rule' for evolution from one level to a band of levels. But in our Cosmological problem we can evolve to yet larger volumes, and in our toy model we describe this by allowing each state $\psi^{(1)}_i$ to have an amplitude $\epsilon$ per unit time to transition into each of a band of states allowed for the yet larger volume $V^{(2)}$. This structure repeats, so that we can evolve to larger and larger volumes $V^{(k)}$. (We keep the same $\epsilon$ and same level spacing $\Delta E$ at each stage for simplicity.) The question in the toy model now is: if we start with $\psi^{(0)}$ (i.e. with volume $V^{(0)}$) at $t=0$, what will be the evolution of the wavefunction? More precisely, if at time $t$ we compute the contribution to $|\psi|^2$ from the states at a given  volume $V^{(k)}$, then at what value of $k$ will this probability peak? 
 
 The problem is a simple quantum mechanical one, and the answer turns out to be
 \be
k_{peak}(t)\approx {2\pi \epsilon^2\over \Delta}t
\label{qwe3}
\ee
for large $t$. In other words, the volume of the Universe is pushed towards larger values $V^{(k)}$. This push is caused by the larger phase space available at larger $k$, and is the Cosmological analog of the evolution of a collapsing shell into the large space of fuzzballs. Thus any other factors causing expansion of the Universe are not included here; (\ref{qwe3}) only describes the expansion generated by consideration of phase space volume. For this effect to be significant it is essential that we be in the `fractional brane gas' phase where the level densities are high -- order black hole level densities -- so that the smallness of the transition amplitude $\epsilon$ can be cancelled by the largeness of $\rho\sim {1\over \Delta E}$. Such a cancellation was the essential mechanism in  black hole evolution, and it appears plausible that it will play a role in Cosmology as well.

To summarize, we may get new `inflationary pushes' in early Universe physics which are not normally accounted for in ordinary treatments of inflation. Such a push to larger volumes may be relevant also to the accelerated expansion that we observe today. The mass within the Cosmological horizon $R_H$ is of the correct order to make a black hole with radius equal to  $R_H$. Thus it is possible that we should think in terms of black hole states when considering the dynamics of matter on these large scales,  and this may again bring us to the large level density $\rho$ and its tendency to push us to larger volumes. 
 
 \section{Conclusion}
 
 The information paradox has forced us to think a  lot about quantum gravity (see \cite{reviews} for some recent reviews). String theory tells us that microstates of black holes are fuzzballs. The study of extremal holes in gravity has suggested that some kind of structure at the horizon may be essential for consistency \cite{marolf}. How the paradox is resolved may also distinguish between different theories of quantum gravity; for example in string theory we seem to get information out in Hawking radiation, while in loop quantum gravity we get slow leakage of information from long lived remnants \cite{smolin}. Extracting the required physics from black holes may be an essential step to understanding singularities in gravity, particularly the big bang.
 
 \section*{Acknowledgements}

I wish to thank all my collaborators who have, over the years, helped put together the ideas discussed here.
I would also like to thank all the members of the black hole community for valuable comments and discussions. This work was supported in part by DOE grant DE-FG02-91ER-40690.

\end{document}